\documentclass[namedreferences]{solarphysics}

\usepackage[hyperref,optionalrh,showbiblabels]{spr-sola-addons} 
\usepackage{graphicx}        
\usepackage{color}           
\usepackage{breakurl}        

\chardef\us=`\_

\begin{document}

\begin{article}
\begin{opening}

\title{Sunspot Observations and Counting at Specola Solare Ticinese in Locarno since 1957}

\author[addressref={aff1}]{\fnm{Sergio}\lnm~{Cortesi}}
\author[addressref={aff1},corref,email={cagnotti@specola.ch}]{\fnm{Marco}\lnm~{Cagnotti}}
\author[addressref={aff2}]{\fnm{Michele}\lnm~{Bianda}}
\author[addressref={aff2}]{\fnm{Renzo}\lnm~{Ramelli}}
\author[addressref={aff1}]{\fnm{Andrea}\lnm~{Manna}}
\address[id=aff1]{Specola Solare Ticinese, Via ai Monti 146, CH-6605 Locarno, Switzerland}
\address[id=aff2]{Istituto Ricerche Solari Locarno, Via Patocchi, CH-6605 Locarno-Monti, Switzerland}

\runningauthor{S. Cortesi et al.}
\runningtitle{Sunspot Counting}

\begin{abstract}

Specola Solare Ticinese is an observatory dedicated to Sunspot Number
counting, which was constructed in 1957 in Locarno, Southern Switzerland, 
as an external observing station of the Zurich Federal Observatory. 
When in 1981  the responsibility of
the determination of the International Sunspot Number was assumed by the Royal
Observatory of Belgium, Specola Solare Ticinese 
was given the role of pilot station,
with the aim of preserving the continuity in the counting method.
We report the observing procedure and counting rules
applied in Locarno.
\end{abstract}
\end{opening}

\section{The Specola Solare Ticinese: A Historical Overview}
\label{S-specola}

\begin{figure}    
  \centerline{
\includegraphics[height=7.8cm]{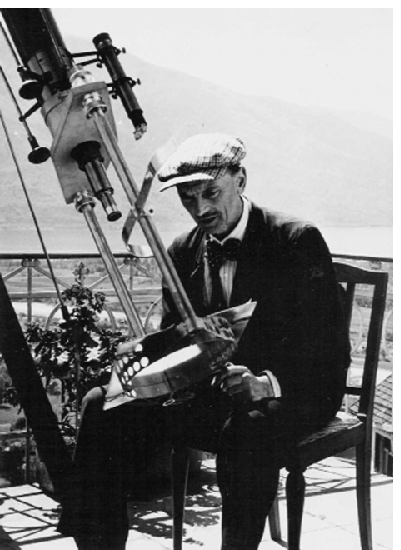}
               \hspace*{0.01\textwidth}
               \includegraphics[height=7.8cm]{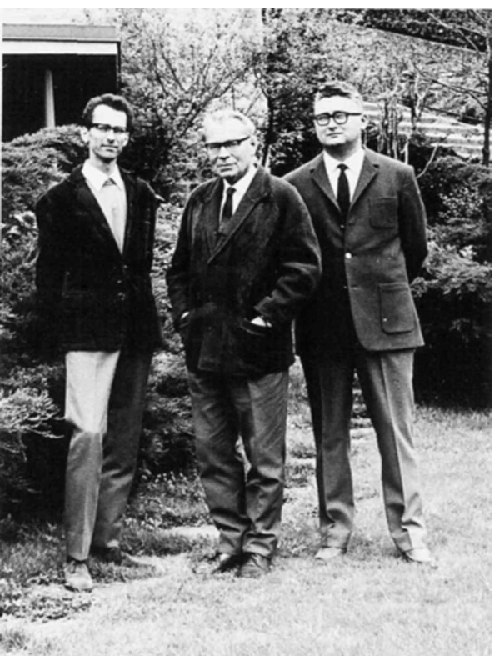}
              }
     \vspace{0.06\textwidth}  
  
\caption{On the left panel, Karl Rapp observing the sun at Locarno with his 14 cm
  aperture Merz refractor. On the right panel, Sergio Cortesi, Max Waldmeier, and
  Araldo Pittini at the entrance of Specola Solare Ticinese, picture taken in
  1961. (Photo credit: S. Cortesi).
        }
   \label{F-Rapp-CPW}
   \end{figure}

\begin{figure}    
  \centerline{\hspace*{0.015\textwidth}
               \includegraphics[width=0.95\textwidth,clip=]{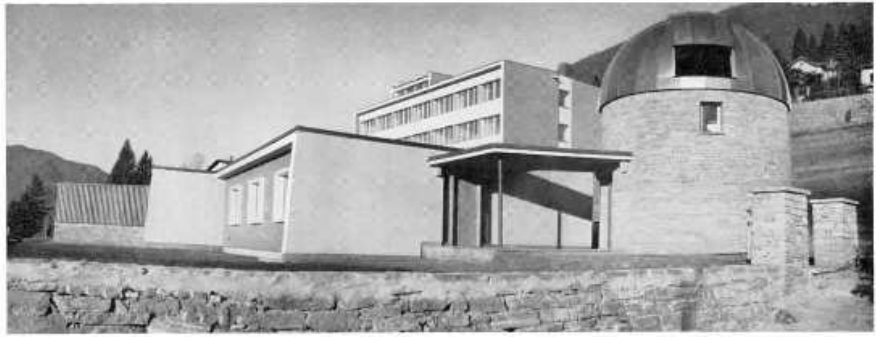}
              }
     \vspace{0.06\textwidth}   

\caption{The Specola Solare observatory just after its
  construction in 1957. In the background is the building of the Swiss
  Meteorological Institute. (Photo credit: S. Cortesi).
    }
   \label{F-specola}
   \end{figure}

\begin{figure}    
  \centerline{\hspace*{0.015\textwidth}
               \includegraphics[height=7cm]{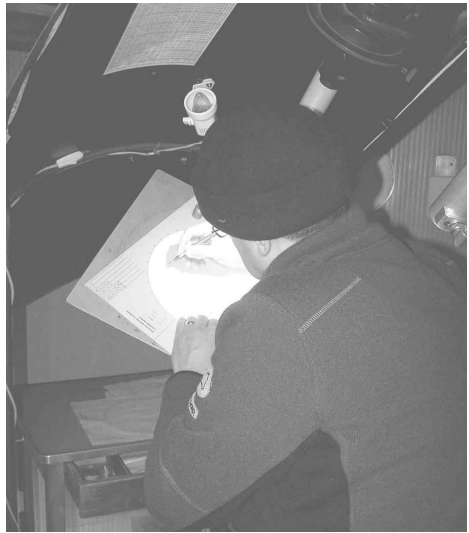}
     \vspace{0.06\textwidth}   
             } 
\caption{Observer Marco Cagnotti when drawing. (Photo credit: S. Cortesi).
        }
   \label{observer}
   \end{figure}

\begin{figure}    
  \centerline{\hspace*{0.015\textwidth}
               \includegraphics[width=0.98\textwidth,clip]{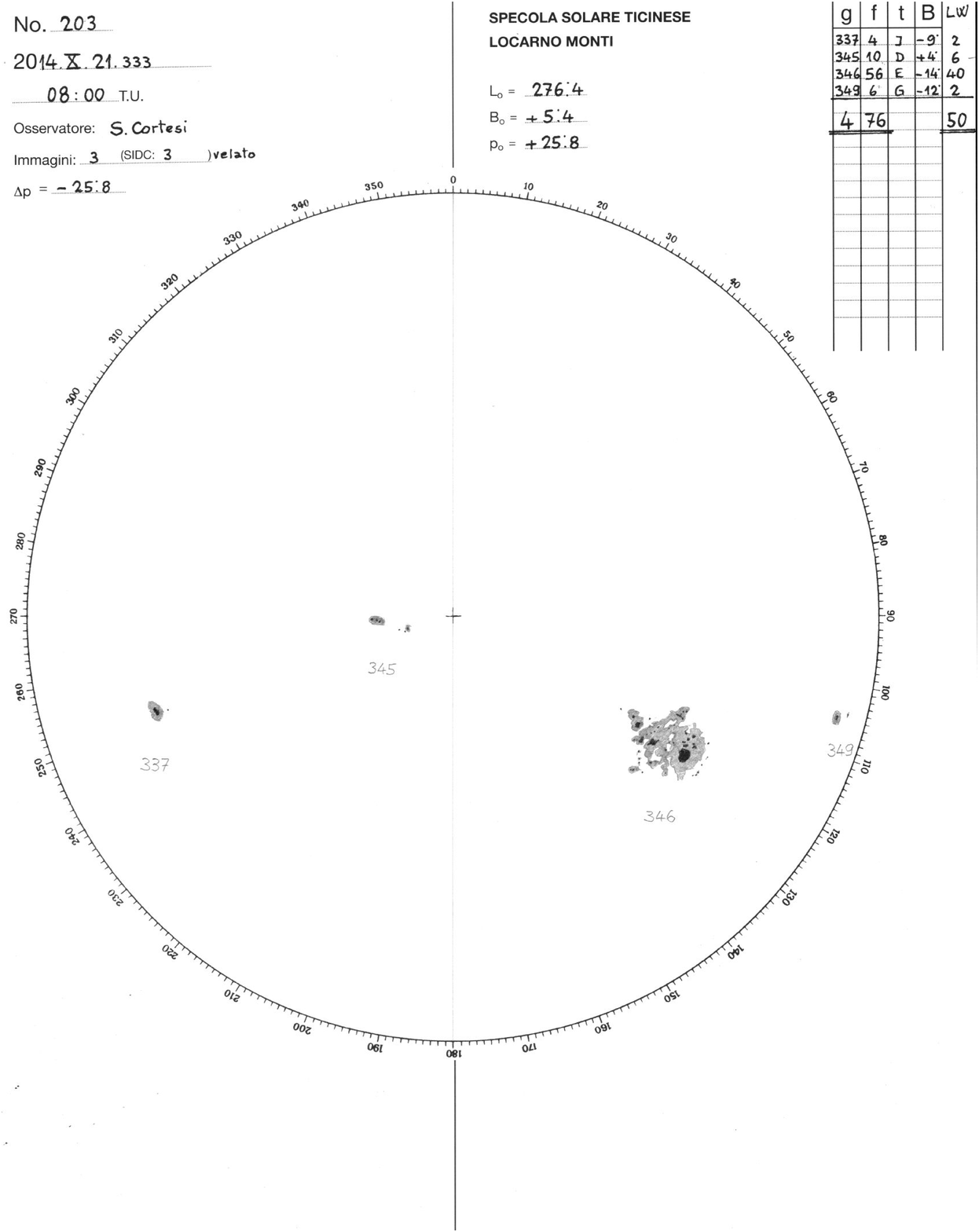}
     \vspace{0.06\textwidth}   
             } 
\caption{Example of a sunspot drawing. 
On top of the drawing sheet the following data are reported: on the
left is the observing information, in the central column are the solar
ephemerides, and on the right table are the group data, including the 
 sequential group number, weighted counting of sunspots, classification,
 latitude, and unweighted counting.
       }
   \label{F-drawing}
\end{figure}

Systematic sunspot observations have been carried out in Locarno, Switzerland,  since the mid-1930s and were started by Karl Rapp
(Figure~\ref{F-Rapp-CPW}, left panel), the engineer who 
founded the Rapp Motorenwerke GmbH in Munich, which afterwards became BMW. Rapp retired in Locarno in 1934 and developed his skills in astronomy
using a 14 cm aperture Merz refractor. His sunspot drawings, sent to William
Brunner at the Zurich Observatory, opened a collaboration that was continued by Max
Waldmeier. The two Zurich researchers 
could appreciate the favorable weather conditions in Locarno
which frequently were complementary on the north and south side of the Alps, due to the meteorological foehn effect. 
Based on this positive experience, Waldmeier decided to build in Locarno the Specola Solare
(Figure~\ref{F-specola}), which was inaugurated in 1957
during the International Geophysical Year. 
The main instrument was the  15 cm aperture Coud\'e--Zeiss refractor. 
This instrument was never changed and is still used now to produce daily  sunspot
drawings based on the projection technique (25cm diameter projected image,
Figure~\ref{observer} and Figure~\ref{F-drawing} ). The first
observers at Specola
Solare were Sergio Cortesi, who already had scientific contacts with Rapp, and
Araldo Pittini (former observer at Zurich Observatory). 

Waldmeier used to spend several weeks per year working in Locarno together
with Cortesi and Pittini, thus having many opportunities 
to instruct them about the counting method and showing them how he was
counting. 
He was pointing out the concept that the main goal for an observer
was to find a constant method of counting. 
He was not in the habit of giving strong rules on how a
single spot had to be taken into account. However the observers,
were invited to check the quality of their work, regularly comparing  their
average counting values with Waldmeier's results.
After an initial learning period, since 1964 the counting method of Locarno's
observers was giving on average the same results as Waldmeier. 
It is worth highlight, that Cortesi
and Pittini developed different personal counting strategies, which
gave different results on a day to day basis, but wich were in agreement on
average over scales of one month or longer.
The observations obtained in Locarno could often fill the gaps due to bad
weather conditions in Zurich.

After  Waldmeier's retirement in 1979,
ETH Zurich decided to stop the sunspot counting activity and to 
 close Specola Solare. The determination of the Wolf number, until
 then always under the responsibility of the Zurich Observatory, 
was assigned by the International Astronomical Union to 
the Sunspot Index Data Center (SIDC) at the Royal Observatory of Belgium. 
The Wolf number was then
renamed  {\it International Relative Sunspot Number}, $R_i$. 
Specola Solare could continue independently his activity managed by
a local association named Associazione Specola Solare
Ticinese. Sergio Cortesi was acting as main observer and gave instructions to
several additional collaborators on the method of counting that
he learned under Waldmeier's guidance.  
From 1981 Specola Solare Ticinese assumed the role of pilot station of SIDC
\citep{clette2014,stenflo2016} in order to minimize the risk of jumps in
transferring the determination of $R_i$ from Zurich to Brussels. 
Since 2012 the role of main observer at Specola Solare Ticinese has been
assumed by Marco Cagnotti after a few years of parallel observations with Cortesi,
and an in--deep instruction on the method transmitted by Waldmeier. When Cagnotti
is unavailable, drawings are still performed by Cortesi 
or other collaborators who have already been active for decades.

\section{Observing and Counting Method} 
      \label{S-counting}      

The  International Relative Sunspot Number is determined with the
formula introduced by \citet{wolf51,wolf56}:
\begin{equation}
R_i = k  (10 N_G + N_S)
\end{equation}
where $N_G$ is the number of groups, $N_S$ the total number of counted 
sunspots and $k$ a scaling factor depending on the instrument and on the observer. 
If weather conditions allow, observations at Specola Solare Ticinese
are performed daily before 12 UT using  
the 15 cm aperture Coud\'e--Zeiss  refractor, 
which is diaphragmed to 8 cm for a better contrast at the specific 
local conditions. 
A projected solar image with 25 cm  diameter is obtained 
on a metallic screen on which a drawing paper sheet 
is fixed, oriented with respect to 
the heliographic axis (see Figures~\ref{observer} and \ref{F-drawing}).  
The observer draws the sunspots with a pencil. The meteorological situation
influences the procedure: in the case of a partially cloudy  sky with short
appearances of the Sun, one is forced to observe quickly, while with a clear
sky, the observer knows that they can take their time. 
In order to distinguish paper sheet
imperfections from real solar structures, when needed
the observer moves a second sheet quickly back and forth above
the drawing sheet. 
This is important in particular to decide
when there is a penumbra, and to find  small sunspots. 
An experienced observer is able to take advantage of a
fraction of a second of good seeing.
It is thus possible to
obtain a drawing corresponding to a photographic image taken with better
seeing quality. 
While the observers are drawing, they are already interpreting the observation,
deciding about the group division and doing the counting of the spots in each
group. 

The original counting method taught by Waldmeier to the Specola observers
provides for different weights according to the sunspot size.

\begin{itemize}
\item a small sunspot is counted as 1 
\item a larger sunspot without penumbra is counted as 2 
\item a small spot with penumbra and only an umbra is counted as 3
\item a big spot with penumbra and only an umbra is counted as 4 or 5
\item a spot with penumbra and 2 umbra is counted as 5
\item more complex sunspots with penumbra and more than two umbrae are counted more than 3, generally the main umbra is counted as 3 and the others are counted as 2 or as 3 depending on the size. 
\item nearby small sunspots can be counted only partially, 
also taking into account the observing conditions: seeing
quality, foggy sky, clear sky, cloudy sky. It can thus happen that the number
of sunspots drawn is larger than the counting.
\end{itemize}

\begin{figure}    
  \centerline{\hspace*{0.015\textwidth}
               \fbox{\includegraphics[height=4cm,clip=]{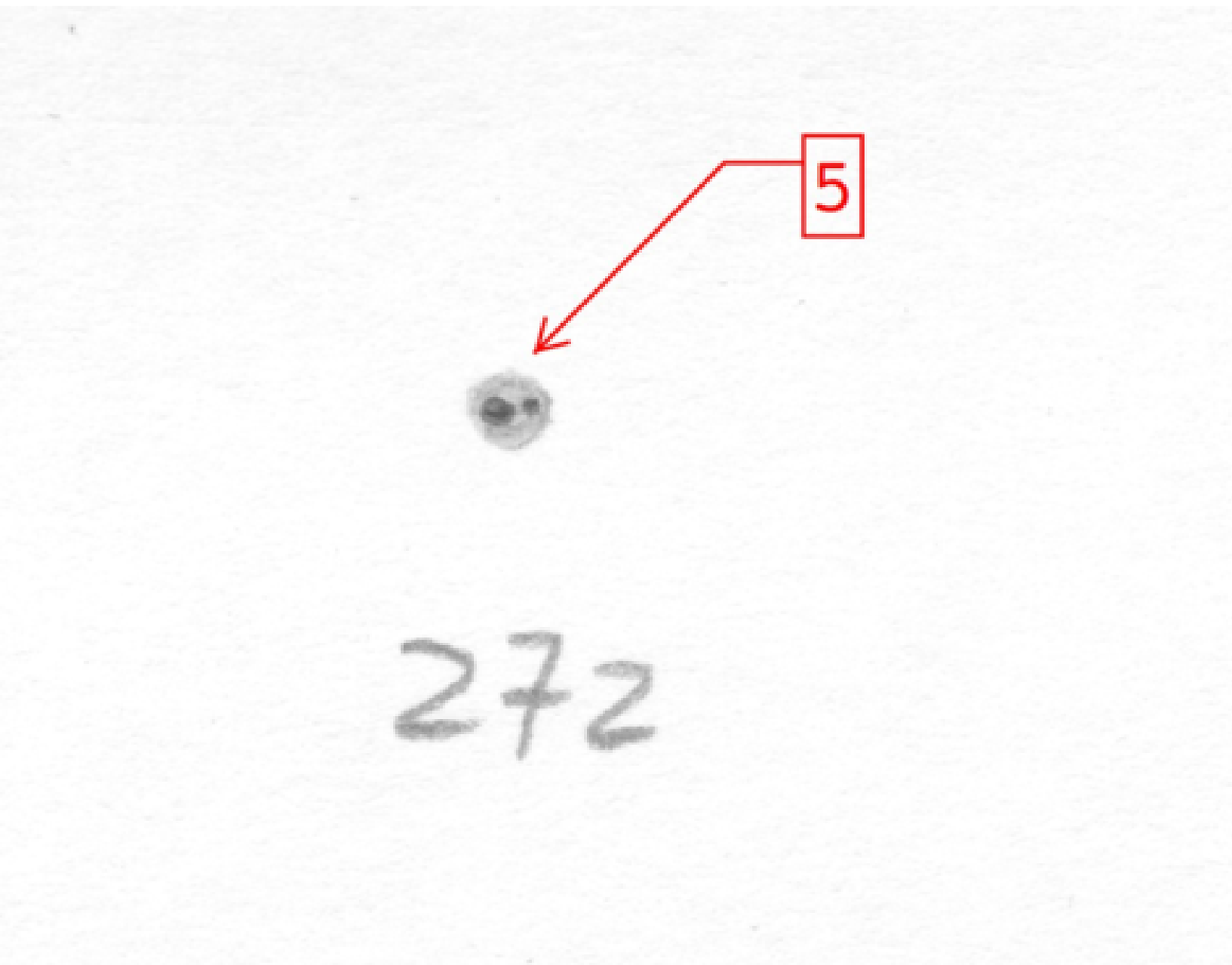}}
              \hspace*{0.03\textwidth}
               \fbox{\includegraphics[height=7cm,clip=]{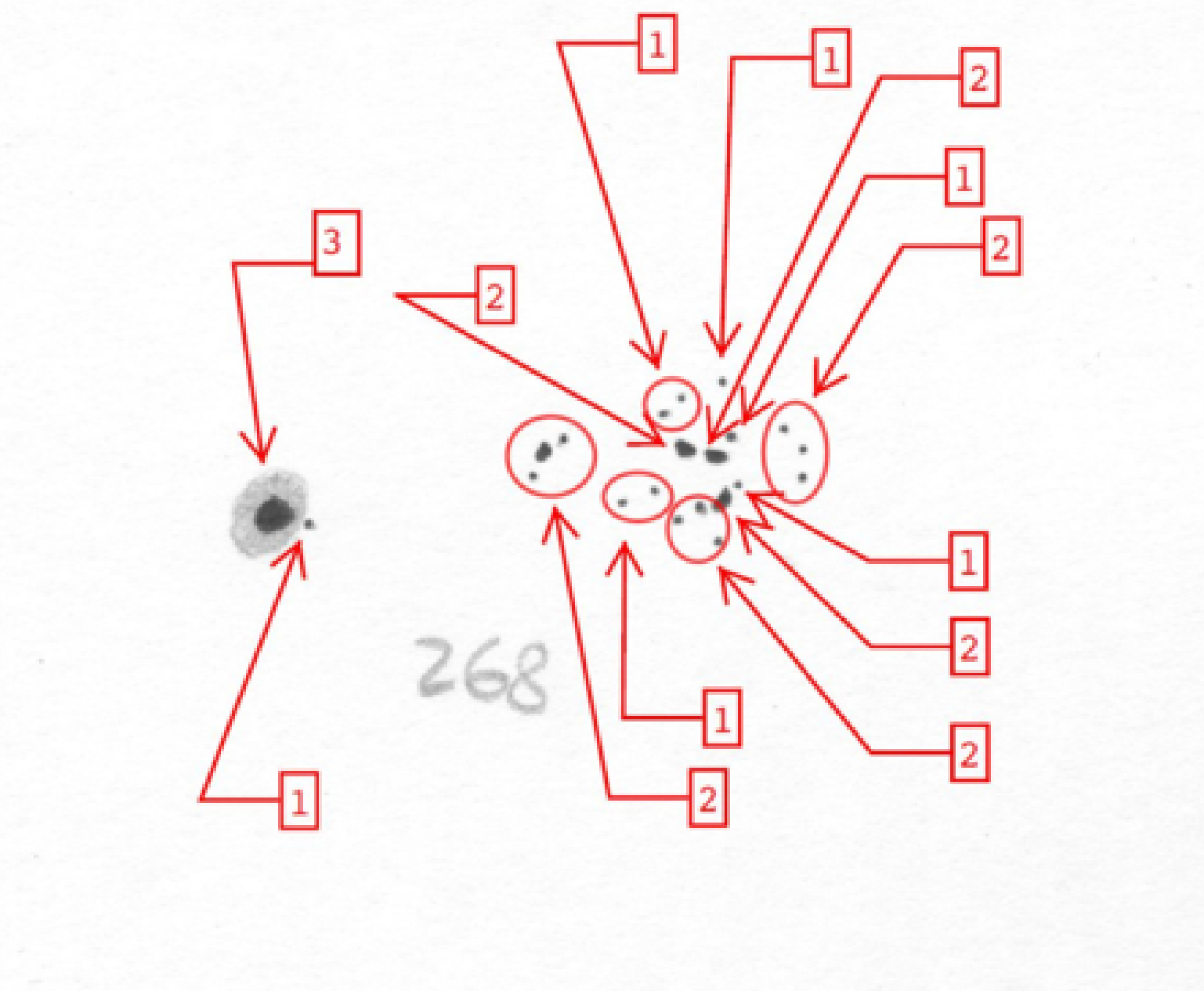}}
     \vspace{0.06\textwidth}   
             } 
\caption{Examples of counting. Here an example on how Cagnotti counted two
  groups reported on the drawing of 26 October 2015.}
   \label{F-cagnotti-counting}
   \end{figure}

Examples of how the weighted counting is applied can be seen in Figure
\ref{F-cagnotti-counting}.

The origin of the weighting method presents some uncertainties. 
The first known written documentation is by \citet{waldmeier48}
\citep[see also][]{waldmeier68}. However
there are some hints that at least
his predecessor Brunner already considered the weighing possibility. In fact
we know that Rapp was already using the weighted counting method based on
Brunner's instructions (private communication of Karl Rapp to Sergio Cortesi
before 1957).

Since 2012 at Specola Solare Ticinese
we started also an unweighted counting method
where each sunspot is counted as 1 independently of the size and of the presence
of a penumbra. In the case of a complex sunspot, with penumbra, each umbra is
counted as 1. This counting is applied later, in the office, counting on the
drawing.
The impact of weighting with respect to the unweighted
counting was studied by \citet{svalgaard15}.

As reported above, the observer also has to decide the grouping of
sunspots. During moderate activity periods this task is in general easy, because
the groups are well separated on the solar disk. However during high activity
periods this could be difficult, because several groups are close to each other
or even partially overlapped. Hints can be given by Joy's law, which
states the group's tendency to be ``tilted'' towards the Equator. 
The Zurich group classification \citep{waldmeier47} is used 
as well, considering the evolution of the groups and looking at the drawings
of the previous days.
It has to be pointed out that in the grouping task 
the observer must avoid taking
advantage of data not available in Wolf's epoch, such as magnetograms or online
high resolution images.

Once the drawing is completed, the observer inserts further information
on the sheet, including the seeing quality in a scale going from 1 (best
quality) to 5 (bad conditions generated by strong air turbulence) and the
observing time with a resolution of a quarter hour.
In addition, the following data of each group are reported in a table:
sequential group number (the numbering is restarted each new year), 
counted number of sunspots (weighted and unweighted), classification, and latitude.

The total and the  hemispheric weighted and unweighted 
counting values are delivered daily to SIDC. 
The drawings are also published online on the Specola web
site (http://www.specola.ch/e/drawings.html).
The archive includes all the drawings since 1981.

Weather conditions in Locarno allows us to obtain on average  306 drawings
per year.

\begin{acks}
The association ASST organizes the financing of Specola Solare Ticinese and
was established by Alessandro Rima (who directed it from 1980 until 1991) and
Paul Utermohlen.  Jan Olof Stenflo at ETHZ recommended using Specola Solare
Ticinese as the pilot station for SIDC in Brussels. Since 1992, ASST has been directed by
Philippe Jetzer. 
Specola Solare Ticinese is mainly financed by Ticino Canton through the
Swisslos found.
We thank Mario Gatti for helpful discussions. 
  \end{acks}


\bibliographystyle{spr-mp-sola}
\bibliography{cortesiV3}

\end{article} 

\end{document}